# QUBODock: A Pip-Installable QUBO Tool for Ligand Pose Generation

Pei-Kun Yang

E-mail: peikun@isu.edu.tw, peikun6416@gmail.com

ORCID: https://orcid.org/0000-0003-1840-6204

## Abstract

We present QUBODock, a pip-installable tool that formulates ligand pose generation as a Quadratic Unconstrained Binary Optimization (QUBO) problem and solves it efficiently on CPU or GPU [1]. QUBODock focuses exclusively on pose generation and deliberately excludes any built-in scoring function, allowing researchers to pair its poses with external scorers of their choice. The software provides a minimal, reproducible interface for (i) protein–ligand structure ingestion and preprocessing, (ii) QUBO model construction from geometric/compatibility constraints, and (iii) decoding solutions into candidate poses for downstream ranking. Implemented in Python with GPU acceleration, QUBODock emphasizes usability and reproducibility: it is distributed on PyPI and can be installed with a single command. We release the source to support benchmarking, teaching, and method development around QUBO-based docking pose generation.

# Introduction

Structure-based virtual screening (SBVS) relies on access to three-dimensional structures of target proteins and candidate ligands [2,3]. In practice, docking is typically used to generate candidate binding poses before any affinity prediction [4-6]. Because large libraries are common and experimentally determined complexes may be unavailable, methods that can produce a small set of geometry-plausible poses efficiently are valuable for downstream scoring and analysis [7].

We present QUBODock, a pose generation-only tool that formulates pose generation as a quadratic unconstrained binary optimization (QUBO) problem and runs on CPU or GPU. QUBODock focuses on generating candidate protein–ligand complex poses and intentionally leaves scoring and reranking to external software chosen by the user. The software is pip-installable, exposes a minimal command-line interface with plain-text inputs and outputs. Although QUBODock ships with a PyTorch-based heuristic QUBO solver for CPU and GPU, the solver layer is modular, and users may substitute external QUBO solvers, including quantum annealers such as D-Wave, digital annealers, or other CPU and GPU optimizers, without changing the pose generation workflow [8-14].

Methodologically, QUBODock adopts a deliberately simple geometric formulation. In typical use, the search region is a spherical pocket defined by a user-provided center and radius, which keeps the grid compact and the number of QUBO variables practical. Within this sphere, we first lay down grid points at a fixed user-specified spacing. We then remove any grid point whose distance to a protein atom is smaller than a preset exclusion radius. Each remaining grid point defines a binary QUBO variable $x_i \in \{0, 1\}$. The objective uses a distance window $[d_{min}, d_{max}]$ over pairs of selected points: pairs with separations inside this window contribute an attractive term that lowers the energy when both variables are 1, pairs closer than $d_{min}$ incur a repulsive penalty, and pairs beyond $d_{max}$ are neutral. This construction encourages solutions in which the $x_i = 1$ points are spread approximately uniformly across the pocket while avoiding steric crowding. The grid spacing controls the size of the candidate set and thus the QUBO, whereas the distance window controls the spatial distribution of the selected points.

From any selected subset of points, QUBODock enumerates rigid ligand placements using distance-consistent atom triplets, computes the corresponding rigid transforms, filters steric clashes against the protein, and, when an experimental pose is available. Inputs are intentionally minimal. A protein .pdb and a ligand .pdb are sufficient, for example, files downloaded from the RCSB PDB [7]. Overall, QUBODock aims for practicality, transparency, and reproducibility as a compact component that supplies pose candidates for SBVS workflows focused on efficient downstream scoring.

## Method

QUBODock generates candidate protein–ligand complex poses by casting pocket-point selection as a QUBO problem, solving it on CPU or GPU, and decoding the selected points into rigid ligand placements. The method is modular: the QUBO can be solved with the built-in PyTorch-based heuristic back end or with external solvers such as quantum annealers and digital annealers, without changing the pose-generation interface [8,10].

**Inputs and preprocessing.** Examples of inputs include a protein .pdb and a ligand .pdb, such as files downloaded from the RCSB PDB [7]. The user specifies a spherical search region by providing a center and radius. This bounds the problem and keeps the number of variables practical on a single GPU or CPU. Within this sphere, QUBODock lays down grid points at a fixed user-specified spacing. Grid points whose distance to any protein atom is smaller than an exclusion distance are removed to avoid steric overlap. Each remaining grid point defines a binary decision variable $x_i \in \{0, 1\}$. Because every grid point maps to one variable, this pruning step directly reduces the size of the QUBO.

**QUBO construction.** Let $G$ denote the set of remaining grid points. QUBODock constructs a QUBO that encourages approximately uniform selections within the spherical region while discouraging local crowding.

$$F(X) = X^\mathrm{T} J X \tag{1}$$

Pairwise couplings are controlled by a distance window $[d_{\min}, d_{\max}]$. For any pair of points $i$ and $j$ whose separation lies inside this window, we set $J_{ij} < 0$ so that selecting both $x_i = 1$ and $x_j = 1$ lowers the energy. If the separation is shorter than $d_{\min}$, we set $J_{ij} > 0$ to penalize co-selection and prevent dense packing. Pairs beyond $d_{\max}$ are treated as neutral ($J_{ij} \approx 0$). The initial grid spacing $s$ determines the size of $G$ and therefore the number of variables, whereas the window $[d_{\min}, d_{\max}]$ shapes the spatial distribution of the points with $x_i = 1$. Larger $s$ yields fewer variables and a smaller QUBO; smaller $s$ increases resolution at the cost of a larger problem.

**Solvers and device selection.** QUBODock provides two built-in solvers, selectable with qubodock-solve --method {sa, greedy}. Either method returns a binary vector that identifies the selected pocket points. Runs can target CPU or CUDA GPUs and default to automatic device selection via --device {auto, cpu, cuda}. The constructed QUBO can also be exported for external engines, such as D-Wave quantum annealers, digital annealers, or other CPU and GPU optimizers, and the resulting binary solution can be re-imported without changing the pose-generation interface [8,10].

**Pose enumeration from selected points.** Given a binary solution $X$, let $S = \{i \in G : x_i = 1\}$. QUBODock enumerates rigid ligand placements by matching distance-consistent atom triplets between S and the ligand. For each ligand atom triplet whose inter-atomic distances are within tolerances of a triplet in $S$, QUBODock computes the rigid transform that superposes the triplets, applies it to the entire ligand, and obtains a candidate pose. This procedure exploits only distances and rigid transforms, which makes it reproducible and insensitive to global superposition choices [1].

**Post-filters and RMSD evaluation.** Each candidate pose is filtered by a steric clash check against protein atoms using a user-selectable clash cutoff. When an experimental pose is available, QUBODock computes RMSD with respect to that pose without any additional superimposition so that candidates can be ranked fairly. When no experimental pose is available, the method reports candidates without internal scoring and leaves ranking to external scorers chosen by the user.

**Implementation Details**

QUBODock is organized as five small command-line programs that exchange plain-text files, making each stage inspectable and replaceable. All coordinates are in Å units. Each program provides sensible defaults; users typically change only a few parameters such as the pocket center and radius, the grid spacing and exclusion radius, solver choice, and matching tolerances.

**qubodock-buildj.** This stage builds the pocket grid and the sparse quadratic couplings that define the QUBO. The pocket is specified by a spherical region using --center X Y Z and --radius, or by enabling --auto-center to estimate a center from the protein heavy atoms. Inside this sphere, grid points are laid down at a fixed spacing controlled by --spacing. To avoid steric overlap, points within an exclusion distance of protein atoms are removed via --exclusion; hydrogens can be included in this pruning with --include-hydrogen. Pairwise couplings are created from a distance window set by --dmin and --dmax, with --reward encouraging co-selection of points whose separation lies inside the window and --penalty discouraging pairs that are too close. Outputs include a sparse edge list of the quadratic form selected by --out (for example j.txt) and the grid-point coordinates selected by --points-out (for example grid_points.txt), preserving the mapping between variables and 3D positions.

**qubodock-solve.** This stage solves the QUBO and returns a binary solution together with the coordinates of the selected points. The solver is chosen with --method {sa, greedy}. Device selection is controlled by --device {auto, cpu, cuda}, and runs can be made reproducible with --seed. For simulated annealing (sa), the temperature schedule and run length are set by --T0, --Tend, and --iters; the optional --density sets the initial fraction of ones in the starting state. For the greedy solver (greedy), --density controls the target fraction of selected bits and thus the stopping condition. The binary solution file is set by --out (for example solution_bits.txt). If the original grid file is provided through --points grid_points.txt, the coordinates of the active points are written with --active-out (for example active_points.txt). Optional timing information can be saved with --time-out. When --device cuda is requested but no compatible GPU is available, the run falls back to the CPU.

**qubodock-align.** This stage decodes a binary selection into rigid ligand placements. It consumes the ligand PDB, the selected pocket points (for example

active_points.txt), and the protein PDB. Matching tolerances are controlled by --pair-tol and --tri-tol (in Å): --pair-tol bounds the allowed deviation between a ligand inter-atomic distance and the corresponding distance between two selected pocket points during seeding and early pruning, whereas --tri-tol requires that all three edges of a candidate atom-triplet match within the specified tolerance before a rigid transform is computed. Smaller tolerances yield fewer but cleaner matches and reduce false positives; larger tolerances increase recall and runtime. Steric filtering is controlled by --clash. The compute device can be chosen with --device {auto, cpu, cuda} and internal search granularity with --search-chunk. Hydrogen atoms can optionally be included as anchors with --anchors-include-h. The list of rigid transforms is written with --placements-out (for example placements.txt), and transformed ligand poses can be saved as PDB files using --save-poses.

**qubodock-rmsd.** When an experimental ligand pose is available, this stage evaluates candidates by RMSD without additional global superposition. It accepts the transform list (placements.txt), a modeled ligand PDB, and the experimental ligand PDB. Atom correspondence can be selected with --match-by {auto, name, index}. Device selection is available via --device {auto, cpu, cuda}. Outputs can be left unsorted with --no-sort, and the result file is set by --out (for example placements_rmsd.txt).

**qubodock-applyrt.** This utility materializes an individual pose by applying a chosen rigid transform from placements.txt to a ligand PDB. The index of the pose is chosen with --pose-index, the compute device with --device {auto, cpu, cuda}, and the output filename with --out (for example pose.pdb).

## Installation

QUBODock is available on PyPI and requires Python ≥ 3.8. Because the built-in solvers are implemented in PyTorch, a CPU or CUDA build of PyTorch must be installed before using qubodock-solve. With PyTorch already installed, install the package with pip install qubodock.

## Conclusion

QUBODock provides a minimal and reproducible path from structures to candidate poses by casting pocket point selection as a QUBO, solving on CPU or GPU, and decoding the selected points into rigid ligand placements. It emphasizes practicality and interoperability: inputs are only a protein PDB and a ligand PDB, installation is a single PyPI command, the command-line interface is small and transparent, and the solver layer is modular so that external QUBO engines can be used without changing the workflow. By omitting an internal scoring function, QUBODock cleanly separates pose generation from affinity prediction and allows researchers to pair its outputs with any downstream scorer.

**Data and Software Availability.** Example inputs and reproduction materials used in this study (sample PDBs, region parameters, and command scripts) are openly available on GitHub at: https://github.com/peikunyang/02_QUBO_Dock. The QUBODock software itself is distributed via PyPI as qubodock and can be installed with pip install qubodock.

**Reference**

1. Yang, P.-K. Ligand Pose Generation via QUBO-Based Hotspot Sampling and Geometric Triplet Matching. *arXiv preprint arXiv:2507.20304* (2025).

2. Maia, E.H.B., Assis, L.C., De Oliveira, T.A., Da Silva, A.M. & Taranto, A.G. Structure-based virtual screening: from classical to artificial intelligence. *Frontiers in chemistry* **8**, 343 (2020).

3. Varela-Rial, A., Majewski, M. & De Fabritiis, G. Structure based virtual screening: Fast and slow. *Wiley Interdisciplinary Reviews: Computational Molecular Science* **12**, e1544 (2022).

4. McNutt, A.T., *et al.* GNINA 1.0: molecular docking with deep learning. *Journal of cheminformatics* **13**, 1-20 (2021).

5. Eberhardt, J., Santos-Martins, D., Tillack, A.F. & Forli, S. AutoDock Vina 1.2.0: New docking methods, expanded force field, and python bindings. *Journal of Chemical Information and Modeling* **61**, 3891-3898 (2021).

6. Allen, W.J., *et al.* DOCK 6: Impact of new features and current docking performance. *Journal of computational chemistry* **36**, 1132-1156 (2015).

7. Burley, S.K., *et al.* RCSB Protein Data Bank (RCSB. org): delivery of experimentally-determined PDB structures alongside one million computed structure models of proteins from artificial intelligence/machine learning. *Nucleic Acids Research* **51**, D488-D508 (2023).

8. Willsch, D., *et al.* Benchmarking Advantage and D-Wave 2000Q quantum annealers with exact cover problems. *Quantum Information Processing* **21**, 141 (2022).

9. Sugisaki, K., Toyota, K., Sato, K., Shiomi, D. & Takui, T. A quantum algorithm for spin chemistry: a Bayesian exchange coupling parameter calculator with broken-symmetry wave functions. *Chemical science* **12**, 2121-2132 (2021).

10. Şeker, O., Tanoumand, N. & Bodur, M. Digital annealer for quadratic unconstrained binary optimization: a comparative performance analysis. *Applied Soft Computing* **127**, 109367 (2022).

11. Woods, B.D., Kochenberger, G. & Punnen, A.P. QUBO Software. in *The Quadratic Unconstrained Binary Optimization Problem* 301-311 (Springer, 2022).

12. Yang, P.-K. Comparative Evaluation of PyTorch, JAX, SciPy, and Neal for Solving QUBO Problems at Scale. *arXiv:2507.17770* (2025).

13. Fujimoto, N. & Nanai, K. Solving QUBO with GPU parallel MOPSO. in *Proceedings of the Genetic and Evolutionary Computation Conference Companion* 1788-1794 (2021).

14. Zaman, M., Tanahashi, K. & Tanaka, S. PyQUBO: Python library for mapping combinatorial optimization problems to QUBO form. *IEEE Transactions on Computers* **71**, 838-850 (2021).